# The promise of spintronics for unconventional computing

Giovanni Finocchio,[1,*] Massimiliano Di Ventra,[2] Kerem Y. Camsari,[3] Karin Everschor-Sitte,[4] Pedram Khalili Amiri,[5] and Zhongming Zeng[6]

[1]Department of Mathematical and Computer Sciences, Physical Sciences and Earth Sciences, University of Messina, Messina, 98166, Italy - Tel. +39.090.3975555

[2]Department of Physics, University of California, San Diego, La Jolla, California 92093, USA, - Tel. +1.858.8226447

[3]School of Electrical and Computer Engineering, Purdue University, West Lafayette, Indiana 47907, USA - Tel. +1.765.4946442

[4]Institut für Physik, Johannes Gutenberg-Universität Mainz, DE-55099 Mainz, Germany - Tel. +49.61313923645

[5]Department of Electrical and Computer Engineering, Northwestern University, Evanston, Illinois 60208, USA - Tel. +1.847.4671035

[6] Key Laboratory of Multifunctional Nanomaterials and Smart Systems,, Suzhou Institute of Nano-tech and Nano-bionics, Chinese Academy of Sciences, Ruoshui Road 398, Suzhou 215123, P. R. China - Tel. +86.0512.62872759

## Abstract

Novel computational paradigms may provide the blueprint to help solving the time and energy limitations that we face with our modern computers, and provide solutions to complex problems more efficiently (with reduced time, power consumption and/or less device footprint) than is currently possible with standard approaches. Spintronics offers a promising basis for the development of efficient devices and unconventional operations for at least three main reasons: (i) the low-power requirements of spin-based devices, i.e., requiring no standby power for operation and the possibility to write information with small dynamic energy dissipation, (ii) the strong nonlinearity, time nonlocality, and/or stochasticity that spintronic devices can exhibit, and (iii) their compatibility with CMOS logic manufacturing processes. At the same time, the high endurance and speed of spintronic devices means that they can be rewritten or reconfigured frequently over the lifetime of a circuit, a feature that is essential in many emerging computing concepts. In this perspective, we will discuss how spintronics may aid in the realization of efficient devices primarily based on magnetic tunnel junctions and how those devices can impact in the development of three unconventional computing paradigms, namely, reservoir computing, probabilistic computing and memcomputing that in our opinion may be used to address some limitations of modern computers, providing a realistic path to intelligent hybrid CMOS-spintronic systems.

## 1. Introduction

Semiconductor electronics have been successful mainly because of three technological characteristics: gain (signal amplification by taking advantage of biasing sources), advantageous signal-to-noise ratio (potential to create a signal well above the noise background), and scalability. However, scalability issues, energy consumption and latency are emerging as severe limitations for the performance of our modern computers which are based on semiconductor electronics. This is because the technological developments of the last decades have built upon the computational paradigms pioneered by Turing and von Neumann in the first half of the last Century, based on the idea that processing of information is done by a unit (e.g., the central processing unit, CPU) that is physically distinct from the one where information is stored. It is then not too difficult to understand that this model leads to an obvious latency and bandwidth limitation in the transfer of information between the processing and memory units, thus creating a bottleneck in the actual execution speed and requiring large amounts of energy to move data.

As a first attempt to a workaround solution, technology has moved in the direction of changing the CPU design, from increasing its speed to increasing the parallelization (multi-core processors). However, this has not addressed the fundamental limitations of present computers and has given life to novel research directions attempting to eventually replace CMOS ("beyond CMOS"), and/or von Neumann architectures. However, such a technology has yet to emerge.

On the other hand, thanks to the recent availability of large amounts of data and computational power (primarily from Graphic Processing Units, GPUs), neuromorphic computing architectures have contributed to the resurgence of artificial intelligence based on deep learning techniques for real-life applications.[1, 2, 3, 4, 5] However, the limitations discussed above also concern existing neuromorphic computing where most of the solutions are software based, and neural networks are still simulated on von Neumann architectures. At the hardware level, mimicking a single synapse/neuron still takes a large number of transistors (large area occupancy) which leads to both integration and energy consumption problems. In addition, the hardware development of the neuronal connectivity is a major limitation due to the 2-dimensional nature of planar CMOS integrated circuit technology. Hence, conventional computing cannot fulfill the increasing requirement of hardware for cognitive and recognition tasks, which cost a large amount of computing time and resources. Despite a lot of research efforts, computational operations based on non-standard (non-Turing or von Neumann) paradigms, i.e., unconventional computing, such as quantum computing, are currently mostly in the research and development stage, and there are still no clear alternatives to CMOS, which remains the stalwart of the semiconductor industry.

Considering all these issues, we believe the time is ripe to identify unconventional approaches that can benefit from a technology that offers advantages such as reduced operation time and/or power consumption, and can still be integrated with CMOS, thus creating a hybrid CMOS-based paradigm of computation. It is the opinion of the authors that spintronics, which exploits the spin degree of freedom of the electron together with its charge, can help to implement new functionalities at the device level that can be integrated at the system level with the mature commercial CMOS technology.

A key element of spintronic technology is the magnetic tunnel junction (MTJ), having as its active element two ferromagnets separated by a thin insulating layer whose resistance depends on the relative orientation of the magnetization vectors in the reference layer and the free layer.[6,7] The magnetization of the free layer can be manipulated by current-induced spin-transfer torque originating from the spin-filter effect of the reference layer magnetization. There have been two main milestones during the development of these devices towards the integration with CMOS technology: the discovery of the large tunneling magnetoresistive (TMR) effects for CoFe/MgO-based MTJ[8] and the interfacial perpendicular anisotropy (IPA) that allows the control of the magnetic anisotropy by film thickness, composition, and by electric fields[9]. Nowadays, a capability for volume manufacturing of MTJs is already set up for memory elements in STT-MRAM[10] (spin-transfer-torque magnetic random access memory), and the major semiconductor foundries have already developed the process to integrate MTJs with state-of-the-art CMOS technology.[11,12] INTEL has demonstrated the integration of STT-MRAM with Intel 22FFL technology in 7.2MB memory array,[11] showing that a voltage with an amplitude of 0.4 V is sufficient to achieve the switching in MTJs. In the same study, INTEL has pointed out the stability of the insulating barrier showing the 12-month time trend of the "Shorting across the MgO barrier", working temperature up to 200°C and thermal stability to magnetic parameters after one hour of exposure at temperatures >400°C. Samsung also described its process to integrate STT-MRAM in a 28-nm FDSOI platform.[13] In addition, simulation frameworks and design tools for hybrid CMOS–spintronic circuits are well established.[14]

In view of the low-power requirements of spin-based devices as well as their compatibility with CMOS manufacturing processes, non-volatile memory, high endurance and speed, it makes sense to look at unconventional computing paradigms that can be realized with spintronic devices. Among all the possible approaches, here we will focus briefly on three physics-based (non-quantum) promising directions: *reservoir computing*,[15] *probabilistic computing*,[16] and *memcomputing*.[17]

Much of the content of this perspective is intended to stimulate the discussion of new experimental approaches and new interdisciplinary research activities at the devices/physics level for which spintronics can play a major role. However, in general, the research efforts should involve synergistic developments at the device-process-circuit-algorithm-system-level.

**2. Spintronics in the real world: From fundamental physics to manufacturing**

**Fundamentals of magnetic tunnel junction based devices.** Research on the development and optimization of MTJ devices is still very active. A key metric is the TMR ratio, defined as ($R_{AP}$ − $R_P$)/$R_P$, where $R_{AP}$ and $R_P$ are the high and low resistance states of the device, respectively. High TMR is important for the read-out of the MTJ state, as a large TMR ratio allows the read-out circuit to reach a particular voltage margin (with respect to a reference value) faster, hence making the read time shorter. In addition, high TMR is an essential requirement for scalability. Given that in a real array of MTJ devices there will be distributions of both $R_{AP}$ and $R_P$ around their mean values, a high TMR ratio ensures sufficient separation of these distributions to perform reliable read-out. Currently, the FeCo/MgO material combination provides the largest TMR ratios (> 200% in perpendicular MTJs with interfacial perpendicular magnetic anisotropy (PMA)) that are being used for embedded applications (i.e. SRAM replacement) where the main limitation is the large current density needed to write at the sub-ns time scale, while replacing higher-capacity standalone memory (e.g. DRAM) may require much higher TMR ratios.

Beyond the standard spin-transfer-torque from a polarized current, two novel promising means for the magnetization manipulation of the free layer are emerging that offer the concomitant reduction of the writing energy and the increase of endurance. The first is the spin-orbit-torque (SOT) in three terminal devices where the MTJ is built on top of a heavy metal with large spin-orbit coupling, and the SOT originates from the current flowing in the heavy metal.[18] Considering SOT-MRAM, the main advantage is the separation of the writing and reading terminals at the cost of smaller bit density as compared to the two-terminal MTJs at the cost of a larger footprint.[19] The other manipulation mechanism is the voltage controlled magnetic anisotropy (VCMA). In VCMA-MRAM, the thickness of the insulating barrier of the MTJ is larger so as to significantly reduce the current flowing into the device and hence the writing power.[20] However, in this case writing is unipolar (i.e., uses only one polarity of voltage) for both directions, and is therefore based on toggling the device to its opposite state using a precisely timed write pulse. Due to the difficulty of controlling the duration of the sub-ns current pulse to avoid increased write errors[21], the most likely path for implementing VCMA-MRAM is to either perform write verification in the CMOS control circuitry,[22] or to combine it with a deterministic write mechanism such as SOT, thus simultaneously achieving the benefits of both approaches. A complete comparison of MTJs with other solutions is beyond the scope of this perspective and we refer the reader to more specialized papers.[7,23]

MTJs have also recently found applications in large-scale neuromorphic computing hardware[2]. In particular, stochastic MTJs, which can be realized by scaling the dimension below the superparamagnetic limit[24] of the free layer or by controlling its effective anisotropy with VCMA,[25] have been used to emulate the functionality of fire-spiking neurons with success. MTJ-based oscillators exhibit large frequency tunability by current and field thanks to the coupling between their power and phase.[26] This property makes them very promising for learning and recognition tasks, in particular vowel recognition has been already demonstrated experimentally with a network of four coupled oscillators.[27] The spintronic diode effect, taking place when an ac current flows thought an MTJ, gives rise to a measurable dc voltage across the MTJ for frequencies near the ferromagnetic resonance.[28] Since its discovery, the performance of resonant spintronic diodes, in terms of sensitivity (output voltage over input power), has been improved significantly and now outperforms the semiconductor counterpart, i.e. Schottky diodes.[29] In fact, biased spintronic diodes have shown sensitivity larger than 200 kV/W[30] and the capability to detect input power below 1 nW.[31] A recent research on mimicking neurons based on spintronic diodes reveals the fact that these tunable devices with both field and current are promising for sparse neuromorphic computing.[32] More information about neuromorphic computing with spintronics can be found in recent

literature.[2,33,34]

**Manufacturing of magnetic tunnel junction based devices.** CMOS technology and its accompanying logic architecture and memory hierarchy have been developed and optimized over several decades, for use in conventional von Neumann computing systems. Fig. 1(a) shows MTJ-based spintronic devices compared against CMOS with memory as the device type of comparison.

The existing memory hierarchy is characterized by a tradeoff between speed (i.e., latency) and cost per bit (i.e., density). This tradeoff has been the determining factor in the choice of various memory elements for use in conventional von Neumann architectures. While some emerging memories (broadly classified as storage-class memories) and some of the early MRAM types are still positioned on the same tradeoff line, a move to unconventional computing requires memories that radically break with the traditional cost-performance paradigm of semiconductors together with an increase in manufacturing maturity, pushing towards the white space (preferred corner) on the top right of Fig. 1a.

Advances in spintronics, with the demonstration of some device properties such as the ones reported in Fig. 1b, have generally pushed magnetic memories towards this direction, with the existing STT-MRAM roughly standing on the same tradeoff line as traditional memories, and new emerging device concepts like SOT and VCMA going beyond it. We wish to highlight here that a key future challenge is to find an experimental strategy to implement spintronic devices that employ all or most of the properties reported in Fig. 1(b) at the same time. From a fundamental point of view, the desired advances in speed and scaling will require both new physics (SOT and VCMA) and new materials, with prime examples being antiferromagnets (AFM) (due to their much faster internal exchange-dominated dynamics)[35], ferrimagnets,[36] topological materials (such as topological insulators),[37] and spin textures (e.g. skyrmions) which may provide better scaling scenarios[38,39]. We believe that the next decade will see an increased dominance of hybrid CMOS-spintronic computing architectures based on STT-MRAM, and subsequently emerging types of MRAM such as SOT and VCMA as they mature technologically.

MRAM is currently being implemented in 28 and 22 nm nodes at various foundries, initially in Fully-Depleted Silicon-on-Insulator (FD-SOI) and subsequently also demonstrated in Fin Field-Effect-Transistor (FinFET) processes, with research and development ongoing for more advanced nodes (14/16 nm and 10 nm) based on STT-MRAM.[11] This already represents a scaling advantage over embedded Flash memory (the main type of memory being displaced initially by embedded STT-MRAM), while also saving cost due to lower mask count. Ultimately, using novel physics (SOT and VCMA) and new materials (e.g. antiferromagnets), scalability to below 7 nm node and higher speed to replace also embedded SRAM may be possible over the next decade.[11,12] In terms of raw performance metrics, MRAM beats all other emerging nonvolatile memories by very significant margins: STT-MRAM has bit-level write energy ~100 fJ/bit,[40,41] with VCMA-MRAM switching demonstrated at < 10 fJ/bit,[42] by far the lowest of any nonvolatile memory device. It has also virtually unlimited endurance, a key requirement for computing applications due to the frequent switching of the device. Endurance can in many cases also be traded off for better speed and retention, if required. It has been demonstrated that VCMA-based MTJs are uniquely suited for integration into cross-point arrays with unidirectional diodes such as Schottky diodes (due to the unipolar nature of VCMA-based switching),[43] and we speculate that this strategy can be used to develop very high density STT-MRAM by using 3D cross-point architecture, as already proposed for STT with specialized two-terminal access devices.[44] To summarize, the main characteristics of spintronic technology based on MTJs have been all demonstrated individually, and some of them have been also reported to occur simultaneously. Despite this, an experimental effort involving mainly materials science and nanofabrication is necessary to have MTJ devices that employ as many properties reported in Fig. 1(b) as possible.

While some of the emerging ideas in unconventional computing such as memcomputing and even probabilistic computing can in principle be realized on CMOS[45,17], these are generally far from being the best hardware implementations for this purpose. Spintronics can offer a more suitable platform for these unconventional computing concepts. Ultimately, sufficient progress towards the preferred corner on the top right of Fig. 1a will enable designers to envision fully spintronic unconventional computing platforms that should be integrated to both scaled CMOS and any hybrid CMOS-spintronic configuration. Let us briefly discuss some of this progress that is unique to spintronic technology.

**Spintronics with skyrmions.** Nontrivial topologically protected spin textures, such as magnetic skyrmions, have received a lot of attention in the past few years due to their unique static and dynamical properties[46,47]. Their first proposed applications are the racetrack memory, skyrmion-based logic gates,[48] microwave devices,[49] and neuromorphic computing.[50] However, it turns out that, to date, the most feasible applications of skyrmions are unconventional, such as token-based Brownian computing[51] and reshuffler devices for probabilistic computing[52] and reservoir computing.[53] While the electrical nucleation and manipulation of a single skyrmion has been already demonstrated,[54] the missing step towards working devices is the development of a read-out scheme that takes advantage of the TMR of MTJs.

**Antiferromagnetic spintronics.** The fundamental properties driving the study of antiferromagnets for applications are their intrinsic high speed and insensitivity to magnetic fields.[35] The efficient electrical manipulation of the antiferromagnetic order with SOT has been already demonstrated showing both bi-stable switching and memory resistive (memristive) behavior.[55] It may also be possible to control antiferromagnets by electric fields for better efficiency.[56] The path toward integration with CMOS is ambitious and would require solving several challenges, such as the scaling of antiferromagnetic devices and a significant improvement in the read-out mechanism. In addition, it should be highlighted that theoretical predictions show that tunable THz oscillators and detectors can be realized considering antiferromagnetic materials and SOT having the additional advantage to work without a bias field.[57,58] This may be a possible direction for the development of compact THz devices for unconventional computing.

**Spintronics with multiferroic and magneto-electric materials.** The integration of multiferroic materials in the field of spintronics seems to be very promising not only for memories, for example magnetoelectric switching being very energy efficient,[60] but also for new scalable energy-efficient logic devices combining magnetoelectricity and spin-orbit coupling[61]. Broadly, multiferroics that have been proposed to date are either single-phase materials, which exhibit intriguing characteristics but allow for only limited degrees of freedom to engineer devices, or heterostructures coupling ferroelectric and ferromagnetic phases (e.g., via strain[62,63,64]), which provide more design flexibility but may present their own integration challenges.

**Magnon spintronics.** The main concept of magnon spintronics is the conversion of information into magnon currents (spin waves) which can propagate without the need to transfer charge, therefore reducing power dissipation.[65] In addition to conventional Von Neumann computing, magnonics offers interesting opportunities for computing with phase, holographic principles, and may serve as an interconnection scheme for reservoir computing based on coupled oscillators.[66] The main challenges are the conversion efficiency between electrical and magnonic domains, i.e., spin wave transduction, where multiferroic and magnetoelectric materials may offer a solution.[67]

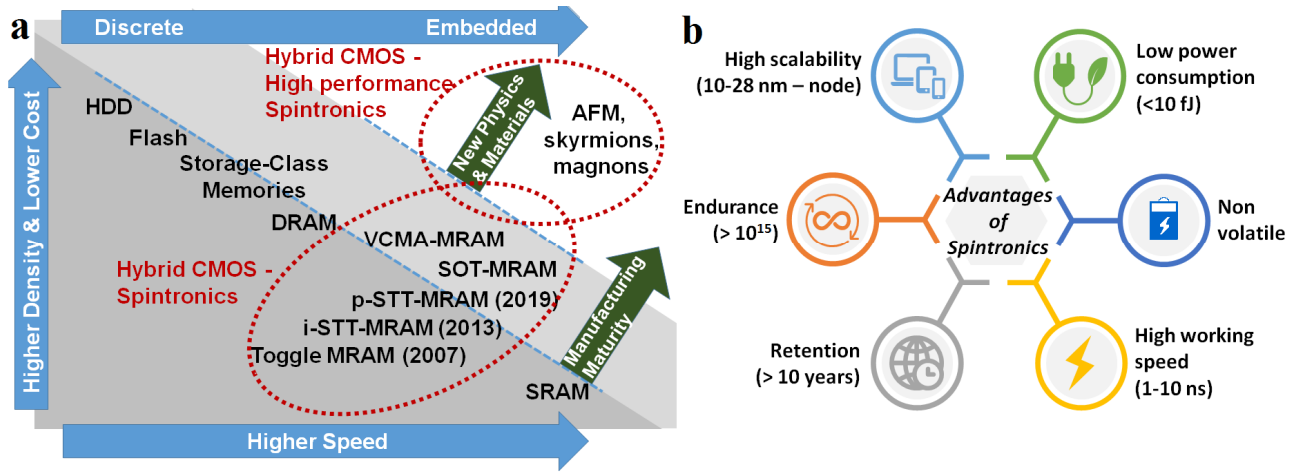

Fig. 1. **a** A competitive landscape of emerging and existing memory technologies, and various spintronic directions within it. The preferred corner is the white space on the top right, which would allow for memories incorporating advanced device concepts such as SOT, VCMA, new materials (e.g. antiferromagnets), and/or spin texture (skyrmions for example). **b** High-level characteristics of MRAM devices integrated with CMOS to date. In particular, integration has been performed in 22 and 28 nm nodes[11] and is under development down to 10 nm based on ferromagnetic devices, while more advanced device concepts and materials will enable scaling to below the 7 nm node.

## 3. Unconventional computing and spintronics

### 3.1 Reservoir computing.

Reservoir computing (RC) is a powerful tool to simplify the classification and separation of spatially and temporally correlated data[68], such as speech recognition, sensor fusion type applications, or nonlinear signal predictions. Originally, RC has emerged from the field of artificial recurrent neural networks (RNNs), which are networks that allow to represent universal Turing machines and general dynamical systems. A RC system consists of an input layer, a reservoir and an output layer. The task of the input layer is to feed the spatial-temporal signal into the system. The complex reservoir then projects the input signal into a sparsely populated higher dimensional space, where the information can be categorized by means of linear regression, see Fig. 2. This directly reveals two of the main advantages of RC: (i) only a small part of the system – the output layer– is trained by a simple, linear regression method and (ii) more than one output can be sampled at the same time. Consequently, the optimal performance of an RC system depends strongly on the properties of the reservoir that can be implemented with any non-linear, complex system with short term memory[69]. There are a plethora of natural systems fulfilling these criteria, opening up the way for efficient in-materio computing, where the material properties (in particular natural memory function, intrinsic complexity and non-linear dynamics) are promoted for computation.[15] We argue that, in the hunt for an ideal reservoir computer, spintronics based systems offer advantages such as low-power consumption, nanoscale and CMOS-compatibility. Several spintronic systems with intricate, non-linear dynamical properties and high tunability have been suggested and shown to work as a reservoir. These include arrays of dipole coupled nanomagnets,[70] spin-wave systems[71], spin-torque oscillators[72] and skyrmion fabrics[73,74], i.e. magnetic textures interpolating between skyrmions, skyrmion lattices and magnetic domain walls. The latter is a highly complex system with a very rich phase space where a random topological magnetic texture (reservoir) in combination with non-linear magneto resistive effects is used to generate complex resistance responses (outputs) to applied voltage patterns injected at nanocontacts (inputs). In such systems the memory is originated in the much slower relaxation of the magnetic texture to a fast-adjusting current profile. The natural excitation frequencies for ferromagnets are in the order of Gigahertz, promising faster inference times compared to state-of-the-art neural network approaches.

Overall, these examples show that spintronic systems offer an excellent perspective to solve complex tasks by means of linear post-processing techniques. It has been shown that operating the reservoir in the region of criticality or near the "edge of chaos" tends to give an optimal performance with respect to the trade-off between complex strong non-linear behavior and sufficiently long memory.[15] As such, current challenges in spintronic based RC include optimizing the reservoirs performance conditions via (externally) tunable parameters, such as applied magnetic fields, finding the best way of injecting the input information, through appropriate preprocessing of the input data as well as improving the signal to noise ratio for room temperature devices providing a fully reliable modus of operation. To speed up the processing, materials with intrinsically faster times scales, such as ferri- and antiferromagnets with natural frequencies reaching up to THz, are a promising direction to explore within the field of in-materio computing.[35]

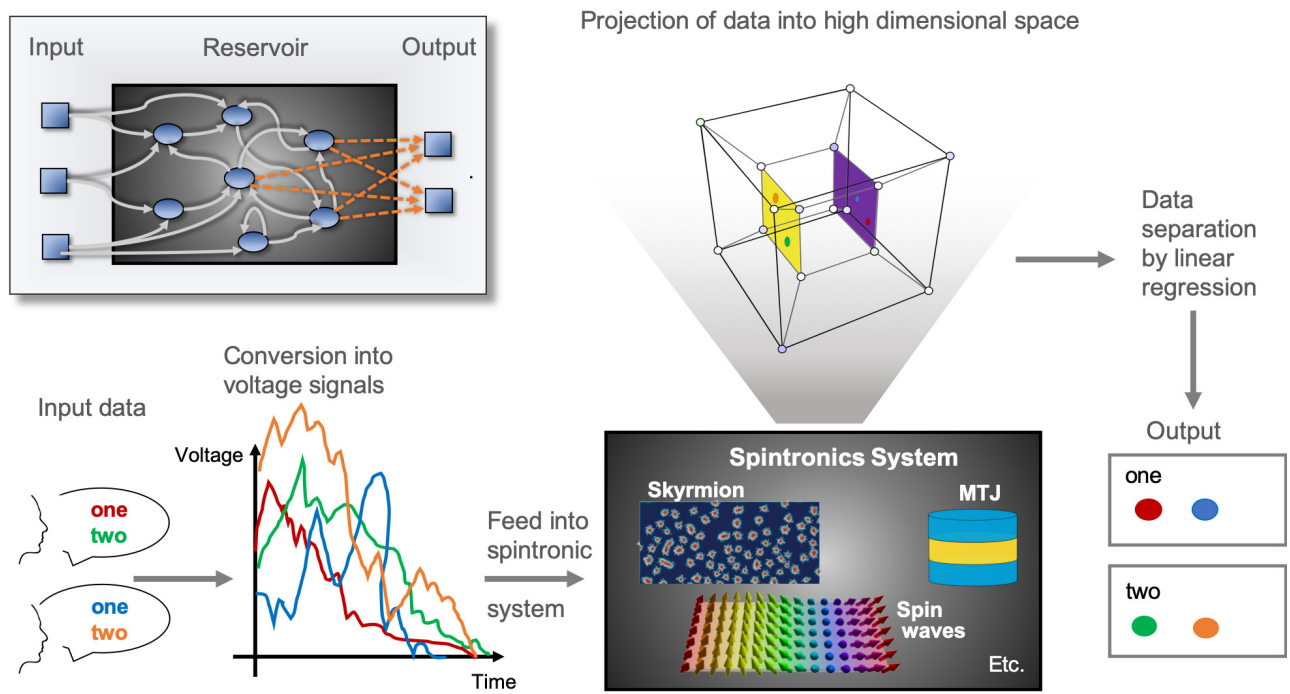

Fig. 2. Schematics for reservoir computing with spintronics. Input data is converted into voltage signals that are fed into the non-linear dynamic spintronics system with a natural memory function, e.g., skyrmion fabric, MTJ or spin-wave system. The system's non-linear dynamics projects the input data into a sparsely populated high-dimensional space, in which data separation can be easily performed by linear regression. Inset on left: reservoir computing scheme based on a random but static recurrent neural network, with feedforward input and output layers. The black box character of these systems indicates that the detailed evolution of the reservoir is not crucial to ensure the functionality of RC, as it relies on only a few basic requirements. The performance of computation however does rely on the properties and quality of the reservoir.

## 3.2 Probabilistic computing.

In the traditional implementation of probabilistic computing, also known as stochastic computing, the information is coded in bit-streams as a probability, p-value (ratio of 1s to the length of the bit-stream), enabling the use of standard logic elements for performing arithmetic operations with p-values. For example, multiplication can be implemented with a simple AND gate.[75] The realization of a skyrmion based reshuffler device, which reshuffles a random bit-stream taking advantage of the Brownian motion of magnetic skyrmions while keeping its p-value constant[52,76] is a first breakthrough for spintronics in this field.

A different paradigm of probabilistic computing is based on the notion of a p-bit that fluctuates between "0" and "1" (Fig. 3(a)).[16] The p-bit is a departure from deterministic bit that is 0 or 1, and a step towards a quantum mechanical qubit that is a coherent superposition of 0 and 1. A three terminal MTJ[18] can be used to generate p-bits (Fig. 3(b)).[77,78] The magnetic energy of its free layer is characterized by two stable states separated by a low barrier, so that thermal fluctuations give rise to fluctuations between those two states, leading to a measurable output voltage through a change in the MTJ resistance. This fluctuating voltage is then thresholded to a binary value by a CMOS-inverter. The probability of the switching can be tuned by the SOT and it is set to be 50% in absence of SOT ($I_{IN}$=0A). In the case of the 1T/1MTJ based three-terminal probabilistic building block (p-bit), the equivalent functionality based on a pseudorandom number generator takes more than 1000 transistors to implement in conventional CMOS.

Probabilistic circuits (p-circuits) that are built out of interconnected p-bits can act as natural hardware units for inherently probabilistic problems such as optimization (for example "Ising Machines") and sampling from a probability distribution.[16] The types of problems that can be addressed by p-circuits are relevant for Machine Learning and Quantum Computing. In the case of Machine Learning, among other applications, p-bits can be used to build energy-efficient "inference" engines, where a network that is trained to recognize a particular type of input is repeatedly used to identify new input data[79]. In the case of Quantum Computing, p-circuits can mimic a subclass of coherent quantum systems to perform quantum annealing, allowing the emulation of quantum annealing using room temperature p-bits.[80] p-circuits can be also designed to build *invertible* logic gates and Boolean circuits that can operate *in reverse*[78] analogous to the concept of self-organizing logic gates and circuits introduced in digital memcomputing machines[81] (see Section 3.3). An experimental proof of concept of integer factorization solved with 8 p-bits generated by MTJs has been presented.[82] Fig. 3(c) shows an example of p-circuit that corresponds to an invertible NAND gate composed of three p-bits A, B and C. This circuit can operate as a usual or invertible NAND, for the latter case the output is clamped to a given value and this causes the p-bits for A and B to fluctuate among the possible consistent alternatives. For example, fixing the output to "1" for a NAND gate makes the inputs fluctuate between (0,1), (1,0) and (0,0) with the same probability.

Several challenges need to be addressed to design scalable p-circuits in hardware. For example, energy barriers of different MTJs could show a significant amount of variation, making each p-bit fluctuate at different time scales. In a symmetrically connected p-circuit, such variations can be tolerated as p-bits can update asynchronously in any random order as long as the synapse network that computes the inputs are faster than p-bits. This requires the use of fast crossbar arrays or CMOS units that operate faster than typical fluctuation rates of nanomagnet based p-bits. In conclusion, a spintronic implementation of probabilistic computing offers two major advantages: (i) the footprint (that it is clearly demonstrated)[82] and energy per random bit is conservatively reduced by 300X and 10X, respectively.[82] (ii) The asynchronous sequential (although effectively parallel) operation can lead to orders of magnitude improvement in convergence time for a given problem.

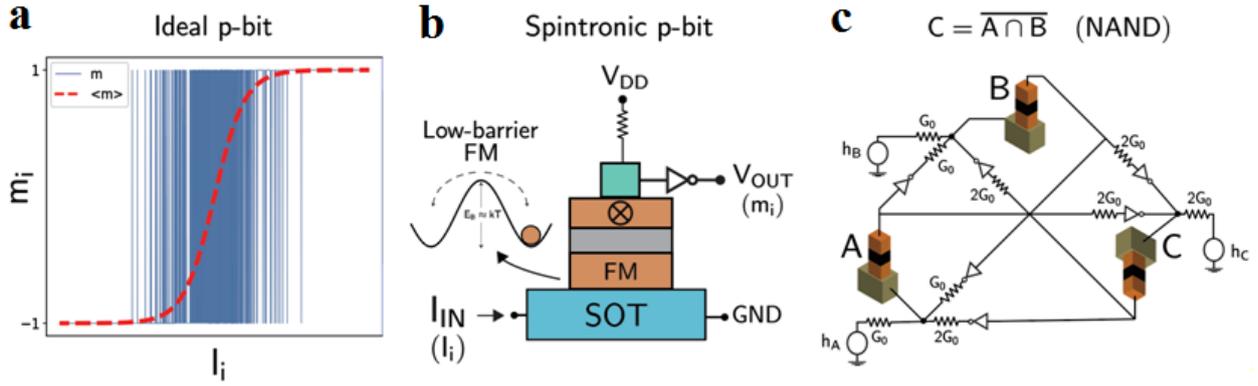

Fig 3. **a** The response of an ideal p-bit, the building block of probabilistic circuits that is mathematically expressed as $m_i$=sgn[tanh($I_i$)-rand(-1,1)][16]. **b** The basic memory element of SOT-MRAM whose free layer is engineered as a low-barrier ferromagnet (FM) with a small thermal stability can function as a spintronic hardware p-bit. **c** An example p-circuit that implements an invertible NAND gate. Unlike standard NAND gates that provide an output (C) for a given set of inputs (A,B), an invertible NAND can find consistent inputs (A,B) for a given output C due to the *invertible* nature of the circuit topology.

**3.3 Memcomputing.**

Memcomputing stands for computing in memory and with memory[17,83,84] and may benefit greatly from a spintronic implementation. Time non-locality (memory) allows for adaptation and self-organization of memcomputing machines to external stimuli, thus facilitating the solution of computationally hard problems efficiently. The digital realization of this paradigm provides a straightforward path to scalable machines. It employs logic gates that are *agnostic* to input and output terminals, namely they can respond to signals coming from the traditional input as well as the traditional output, and dynamically adapt to these signals so as to always satisfy the logic truth table they are meant to represent.[81] In other words, these "self-organizing gates" can operate *in reverse*, as those suggested in probabilistic computing. However, unlike probabilistic computing, the gates in memcomputing are *deterministic*, and as such, the digital memcomputing machines built from them are more easily scalable. In addition, these machine employ topological objects (instantons) in the phase space to reach the solution.[17] Therefore, they are robust against noise and disorder.

In order to accomplish this feat, self-organizing gates have internal degrees of freedom (memory) that allow the gates to go through a continuous dynamics during which the external terminals of the gates are not constrained to be integers. Fig. 4a shows a schematic of these gates arranged into a circuit representing some logic proposition. The internal memory variables can be realized in practice in various ways. For instance, by employing resistive memories (see schematic in Fig. 4b), whose conductance depends on internal degrees of freedom, ($\tilde{x}_j$) (e.g., magnetization), and the voltage (or current): $g_M(x,V)$. These resistive memories, in turn, can be realized using spintronic components.[2,34] Fig. 4c shows an example of domain wall based memristor where the number of resistance states is defined by the free layer geometry of the MTJ and can be controlled by an electric current or magnetic field.[85,86] Recently, is has been also shown that antiferromagnets exhibit memristive behavior.[35] This fact can be important for the realization of ultrafast memristors that may push the memcomputing machines to operate in the THz range.[87] Since these digital machines are non-quantum dynamical systems, their equations of motion can be integrated numerically very efficiently. In fact, several studies, ranging from maximum satisfiability to quadratic unconstrained binary optimization to linear integer programming have already shown that the simulations of these machines offer great advantages compared to traditional algorithms.[17,81] For instance, two unsolved problems of the MIPLIB library (the Mixed Integer Programming Library) have been recently solved with the memcomputing paradigm, f2000[88] and pythago7824[89].

We therefore expect that a hardware implementation of the same machines would provide even greater benefits and a realistic path towards real-time computation with impact on a wide variety of applications, such as autonomous vehicles, robotics, etc. In this respect, the features we have discussed about spin-based devices, in particular, their low power and compatibility with CMOS manufacturing processes, are ideally suited for the hardware implementation of this computing paradigm.

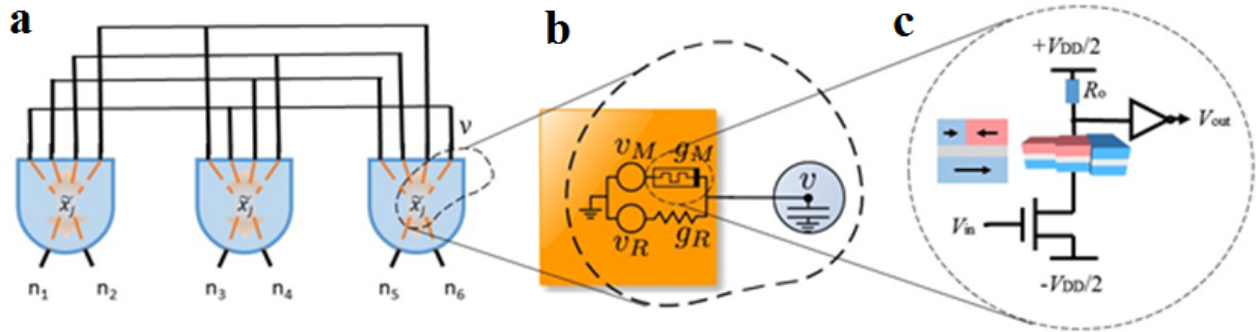

Fig. 4. **a** Schematic of self-organizing gates with internal memory variables, $\tilde{x}_j$, connected to represent a Boolean logic circuit, where the inputs, $n_i$, are assigned logical variables and the voltages, $v$, represent the logical variables that satisfy the Boolean formula. **b** Schematic of the interior of a self-organizing gate realized by resistive memories with conductance $g_M$, standard resistors with conductance, $g_R$, and voltage generators. **c** Schematic of a spintronic memrisistive memory.[85]

## 4. Impact and future directions

By taking advantage of the mature commercial technologies of MTJs, and the immense investment and know-how in related materials, processes, and devices, it is well within reach to implement new computational paradigms integrating spintronic devices with technologies that are ready or close to volume manufacturing. This may initially take the form of heterogeneous integration, where systems on chip and in package may be realized by combining spintronics, CMOS, and other components such as optoelectronics and sensors. The spintronics elements can be integrated monolithically, towards fully integrated spintronics computing platforms with combined memory, logic, and sensing using magnetic materials. Spintronic hardware for unconventional computing can be also implemented monolithically in CMOS logic processes, e.g., to complement traditional CMOS-based computing (see Fig. 5). An example could be probabilistic or memcomputing based co-processors for optimization purposes, or a reservoir computing co-processor implemented to reduce the device footprint of neural networks on chip. In addition, using heterogeneous integration techniques such as chip-level bonding and/or chiplets, one can imagine combinations of these technologies with traditional von Neumann components (e.g., DRAM), sensors and optoelectronics, as well as other emerging devices such as resistive memory circuits.

These examples show that, in the long run, our notion of what computation means will change. In particular, we anticipate that exploiting the nonlinear dynamics of a system will by far be more efficient than simulating artificial neuromorphic entities on conventional transistor-based technology. In particular, we expect that for some applications the trend to be going back from digital to analog information encoding or a mix of the two. Finding novel hardware solutions for unconventional computing that are compatible with CMOS manufacturing technology will be the first step to enter the market towards eventually realizing computational schemes beyond von Neumann computers. As pointed out in this perspective, besides the basic requirements, spintronic-based systems offer small and energy efficient solutions for unconventional computing with natural memory and stochastic behavior.

We conclude by noting that the field of computing is now moving towards the realization of what is referred to as the "third wave" of Artificial Intelligence, namely the creation of computing platforms that boast human-level learning, adapt to the environment, understand and can communicate their limitations, and can tackle problem solving on their own. Although it is not clear when this wave will come crushing on us, it is the opinion of these authors that it can only be implemented with alternatives to our present paradigms and architectures. The unconventional computing paradigms we have discussed here, and their spintronic realizations, offer a first, realistic step towards riding this third wave. We thus hope our perspective will inspire and motivate much-needed research in this fascinating area with far-reaching impact on many fields.

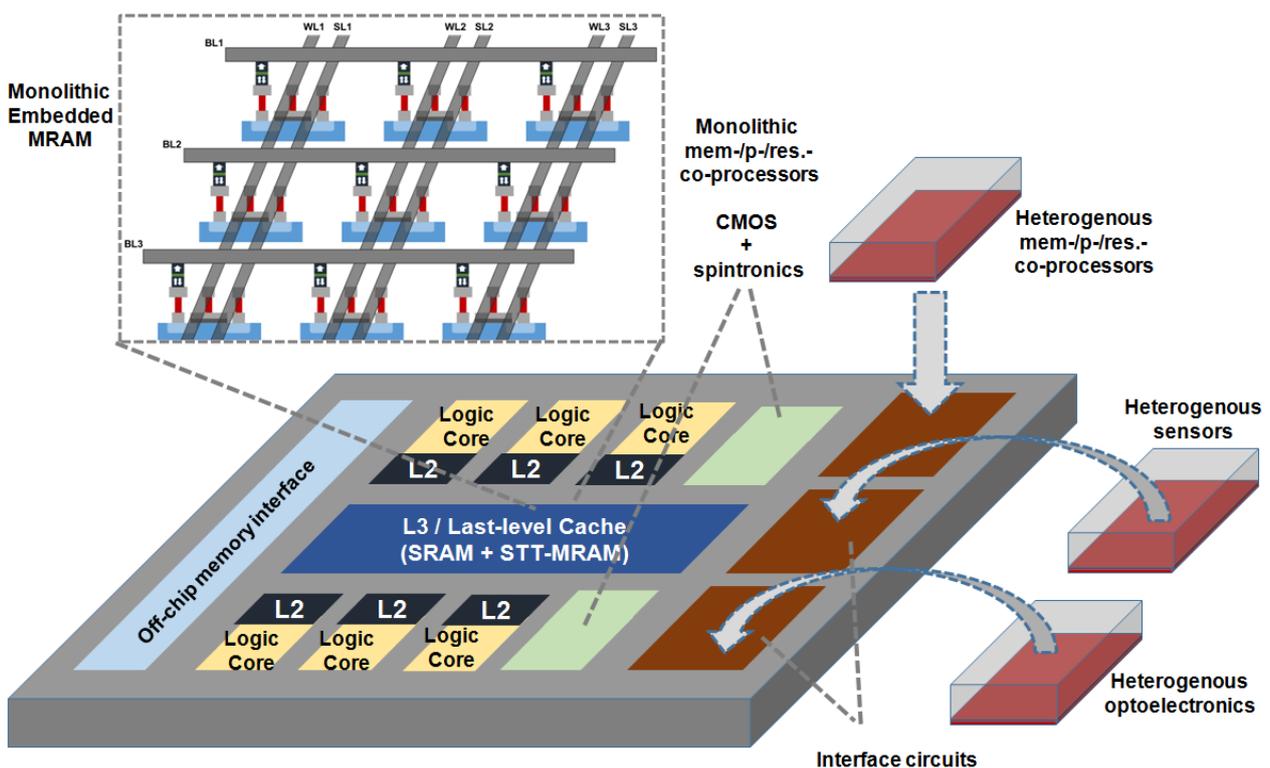

Fig. 5. Illustration of a system on chip combining traditional von Neumann computing architecture with new unconventional computing co-processors. Spintronics is at the heart of the implementation of both computing types on the chip: It is monolithically integrated on CMOS (top left) as a back-end of line process within the metallization layers of the underlying CMOS process (which can be FinFET, FD-SOI, or conventional CMOS). Within the von Neumann paradigm, spintronics will multiply the amount of available on-chip memory (initially, L3 and/or Last-level Cache) by replacing the existing SRAM, which requires 6 to 8 transistors per cell and is area-inefficient. At the same time, this monolithic integration can be used to implement hybrid CMOS plus magnetic tunnel junction circuitry for mem-, p-, or reservoir co-processors. The latter can, optionally, also be integrated heterogeneously if it involves other types of devices not available for monolithic integration in the same CMOS process, e.g., memristors, phase change, or 2D material based devices. Heterogenous integration also allows for integration of other components of the system such as optoelectronics and sensors.

**Acknowledgements**
GF and ZZ acknowledge the Executive Programme of Scientific and Technological Cooperation between Italy and China for the years 2016‑2018 (code CN16GR09, 2016YFE0104100) titled Nanoscale broadband spin-transfer-torque microwave detector. MD acknowledges partial support from the Center for Memory and Recording Research at the UCSD and DARPA (grant HR00111990069). KYC acknowledges useful discussions with Supriyo Datta. KES acknowledges the funding from the German Research Foundation (DFG) under the Project No. EV 196/2-1 and from the Carl-Zeiss-Stiftung through the JGU Research Center for Emergent Algorithmic Intelligence. PKA acknowledges support by a grant from the National Science Foundation, Division of Electrical, Communications and Cyber Systems (NSF ECCS-1853879). GF, MD, PKA and ZZ also acknowledge the support from the school of excellence program "Brain Inspired Computing" at University of Messina and discussions with Prof. Michele Gaeta.

**Contributions**
G.F. drafted the initial article, G.F., P.K.A., Z.Z. wrote the sections on spintronics (Figs. 1 and 5). M.D. wrote the memcomputing section (Fig. 4), K.Y.C wrote the probabilistic computing section (Fig. 3), and K.E.S wrote the reservoir computing section (Fig. 2). All authors reviewed and contributed to the final version of the article.

**Competing interests**
MD is the co-founder of MemComputing, Inc. (https://memcpu.com/) that is attempting to commercialize the memcomputing technology. All other authors declare no competing interests.

**References**

1. Chen, Y. *et al.* Neuromorphic computing's yesterday, today, and tomorrow – an evolutional view. *Integration* **61**, 49–61 (2018).

2. Grollier, J., Querlioz, D. & Stiles, M. D. Spintronic Nanodevices for Bioinspired Computing. *Proc. IEEE* **104**, 2024–2039 (2016).

3. Burr, G. W. *et al.* Neuromorphic computing using non-volatile memory. *Adv. Phys. X* **2**, 89–124 (2017).

4. Sengupta, A. & Roy, K. Neuromorphic computing enabled by physics of electron spins: Prospects and perspectives. *Appl. Phys. Express* **11**, 030101 (2018).

5. Boybat, I. *et al.* Neuromorphic computing with multi-memristive synapses. *Nat. Commun.* **9**, 2514 (2018).

6. Zhu, J.-G. & Park, C. Magnetic tunnel Fueled by the ever-increasing demand for larger hard disk drive storage. *Mater. Today* **9**, 36–45 (2006).

7. Endoh, T. & Honjo, H. A Recent Progress of Spintronics Devices for Integrated Circuit Applications. *J. low power Electron. Appl.* **8**, 1–17 (2018).

8. Parkin, S. S. P. *et al.* Giant tunnelling magnetoresistance at room temperature with MgO (100) tunnel barriers. *Nat. Mater.* **3**, 862–867 (2004).

9. Ikeda, S. *et al.* A perpendicular-anisotropy CoFeB-MgO magnetic tunnel junction. *Nat. Mater.* **9**, 721–724 (2010).

10. Kent, A. D. & Worledge, D. C. A new spin on magnetic memories. *Nat. Nanotechnol.* **10**, 187–191 (2015).

11. Golonzka, O. *et al.* MRAM as Embedded Non-Volatile Memory Solution for 22FFL FinFET Technology. in *2018 IEEE International Electron Devices Meeting (IEDM)* 18.1.1-18.1.4 (IEEE, 2018). doi:10.1109/IEDM.2018.8614620

12. Raychowdhury, A. MRAM and FinFETs team up. *Nature Electronics* **1**, 618–619 (2018).

13. Song, Y. J. *et al.* Demonstration of Highly Manufacturable STT-MRAM Embedded in 28nm Logic. in *International Electron Devices Meeting* talk 18.2 (2018).

14. De Rose, R. *et al.* Variability-Aware Analysis of Hybrid MTJ/CMOS Circuits by a Micromagnetic-Based Simulation Framework. *IEEE Trans. Nanotechnol.* **16**, 160–168 (2017).

15. Tanaka, G. *et al.* Recent advances in physical reservoir computing: A review. *Neural Networks* **115**, 100–123 (2019).

16. Camsari, K. Y., Sutton, B. M. & Datta, S. p-bits for probabilistic spin logic. *Appl. Phys. Rev.* **6**, 011305 (2019).

17. Di Ventra, M. & Traversa, F. L. Perspective: Memcomputing: Leveraging memory and physics to compute efficiently. *J. Appl. Phys.* **123**, 180901 (2018).

18. Liu, L. *et al.* Spin-torque switching with the giant spin hall effect of tantalum. *Science (80-. ).* **336**, 555–558 (2012).

19. Prenat, G. *et al.* Ultra-Fast and High-Reliability SOT-MRAM: From Cache Replacement to Normally-Off Computing. *IEEE Trans. Multi-Scale Comput. Syst.* **2**, 49–60 (2016).

20. Khalili Amiri, P. & Wang, K. L. Voltage-Controlled Magnetic Anisotropy in Spintronic Devices. *SPIN* **02**, 1240002 (2012).

21. Khalili Amiri, P. *et al.* Electric-Field-Controlled Magnetoelectric RAM: Progress, Challenges, and Scaling. *IEEE Trans. Magn.* **51**, 1–7 (2015).

22. Grezes, C. *et al.* Write Error Rate and Read Disturbance in Electric-Field-Controlled Magnetic Random-Access Memory. *IEEE Magn. Lett.* **8**, 1–5 (2017).

23. Dieny, B. *et al.* Opportunities and challenges for spintronics in the microelectronic industry. (2019).

24. Vodenicarevic, D. *et al.* Low-Energy Truly Random Number Generation with Superparamagnetic Tunnel Junctions for Unconventional Computing. *Phys. Rev. Appl.* **8**, 1–9 (2017).

25. Cai, J. *et al.* Voltage-Controlled Spintronic Stochastic Neuron Based on a Magnetic Tunnel Junction. *Phys. Rev. Appl.* **11**, 034015 (2019).

26. Slavin, A. & Tiberkevich, V. Nonlinear Auto-Oscillator Theory of Microwave Generation by Spin-Polarized Current. *IEEE Trans. Magn.* **45**, 1875–1918 (2009).

27. Romera, M. *et al.* Vowel recognition with four coupled spin-torque nano-oscillators. *Nature* **563**, 230–234 (2018).

28. Tulapurkar, A. A. *et al.* Spin-torque diode effect in magnetic tunnel junctions. *Nature* **438**, 339–342 (2005).

29. Miwa, S. *et al.* Highly sensitive nanoscale spin-torque diode. *Nat. Mater.* **13**, 50–56 (2014).

30. Zhang, L. *et al.* Ultrahigh detection sensitivity exceeding 10 5 V/W in spin-torque diode. *Appl. Phys. Lett.* **113**, 102401 (2018).

31. Fang, B. *et al.* Experimental Demonstration of Spintronic Broadband Microwave Detectors and Their Capability for Powering Nanodevices. *Phys. Rev. Appl.* **11**, 014022 (2019).

32. Cai, J. *et al.* Sparse neuromorphic computing based on spin-torque diodes. *Appl. Phys. Lett.* **114**, 192402 (2019).

33. Zhirnov, V. V. *et al.* Memory Devices: Energy–Space–Time Tradeoffs. *Proc. IEEE* **98**, 2185–2200 (2010).

34. Grollier, J., Guha, S., Ohno, H. & Schuller, I. K. Preface to Special Topic: New Physics and Materials for Neuromorphic Computation. *J. Appl. Phys.* **124**, 151801 (2018).

35. Jungwirth, T., Marti, X., Wadley, P. & Wunderlich, J. Antiferromagnetic spintronics. *Nat. Nanotechnol.* **11**, 231–241 (2016).

36. Finley, J. & Liu, L. Spin-Orbit-Torque Efficiency in Compensated Ferrimagnetic Cobalt-Terbium Alloys. *Phys. Rev. Appl.* **6**, 054001 (2016).

37. Mellnik, A. R. *et al.* Spin-transfer torque generated by a topological insulator. *Nature* **511**, 449–451 (2014).

38. Parkin, S. & Yang, S.-H. Memory on the racetrack. *Nat. Nanotechnol.* **10**, 195–198 (2015).

39. Fert, A., Cros, V. & Sampaio, J. Skyrmions on the track. *Nat. Nanotechnol.* **8**, 152–156 (2013).

40. Khalili Amiri, P. *et al.* Low write-energy magnetic tunnel junctions for high-speed spin-transfer-torque MRAM. *IEEE Electron Device Lett.* **32**, 57–59 (2011).

41. Zhao, H. *et al.* Low writing energy and sub nanosecond spin torque transfer switching of in-plane magnetic tunnel junction for spin torque transfer random access memory. *J. Appl. Phys.* **109**, 1–4 (2011).

42. Grezes, C. *et al.* Ultra-low switching energy and scaling in electric-field-controlled nanoscale magnetic tunnel junctions with high resistance-area product. *Appl. Phys. Lett.* **108**, 012403 (2016).

43. Dorrance, R. *et al.* Diode-MTJ crossbar memory cell using voltage-induced unipolar switching for high-density MRAM. *IEEE Electron Device Lett.* **34**, 753–755 (2013).

44. Yang, H. *et al.* Threshold switching selector and 1S1R integration development for 3D cross-point STT-MRAM. in *2017 IEEE International Electron Devices Meeting (IEDM)* 38.1.1-38.1.4 (IEEE, 2017). doi:10.1109/IEDM.2017.8268513

45. Smithson, S. C., Onizawa, N., Meyer, B. H., Gross, W. J. & Hanyu, T. Efficient CMOS invertible logic using stochastic computing. *IEEE Trans. Circuits Syst. I Regul. Pap.* **66**, 2263–2274 (2019).

46. Finocchio, G., Büttner, F., Tomasello, R., Carpentieri, M. & Kläui, M. Magnetic skyrmions: from fundamental to applications. *J. Phys. D. Appl. Phys.* **49**, 423001 (2016).

47. Everschor-Sitte, K., Masell, J., Reeve, R. M. & Kläui, M. Perspective: Magnetic skyrmions—Overview of recent progress in an active research field. *J. Appl. Phys.* **124**, 240901 (2018).

48. Zhang, X., Ezawa, M. & Zhou, Y. Magnetic skyrmion logic gates: conversion, duplication and merging of skyrmions. *Sci. Rep.* **5**, 9400 (2015).

49. Finocchio, G. *et al.* Skyrmion based microwave detectors and harvesting. *Appl. Phys. Lett.* **107**, 262401 (2015).

50. Li, S. *et al.* Emerging Neuromorphic Computing Paradigms Exploring Magnetic Skyrmions. in *2018 IEEE Computer Society Annual Symposium on VLSI (ISVLSI)* 539–544 (IEEE, 2018). doi:10.1109/ISVLSI.2018.00104

51. Nozaki, T. *et al.* Brownian motion of skyrmion bubbles and its control by voltage applications. *Appl. Phys. Lett.* **114**, 012402 (2019).

52. Zázvorka, J. *et al.* Thermal skyrmion diffusion used in a reshuffler device. *Nature Nanotechnology* 4–9 (2019). doi:10.1038/s41565-019-0436-8

53. Pinna, D., Bourianoff, G. & Everschor-Sitte, K. Reservoir Computing with Random Skyrmion Textures. *arXiv 1811.12623* (2018).

54. Büttner, F. *et al.* Field-free deterministic ultrafast creation of magnetic skyrmions by spin-orbit torques. *Nat. Nanotechnol.* **12**, 1040–1044 (2017).

55. Wadley, P. *et al.* Electrical switching of an antiferromagnet. *Science (80-. ).* **351**, 6273 (2016).

56. Lopez-Dominguez, V., Almasi, H. & Amiri, P. K. Picosecond Electric-Field-Induced Switching of Antiferromagnets. *Phys. Rev. Appl.* **11**, 024019 (2019).

57. Gomonay, E. V. & Loktev, V. M. Spintronics of antiferromagnetic systems (Review Article). *Low Temp. Phys.* **40**, 17–35 (2014).

58. Puliafito, V. *et al.* Micromagnetic modeling of terahertz oscillations in an antiferromagnetic material driven by the spin Hall effect. *Phys. Rev. B* **99**, 1–7 (2019).

59. Khymyn, R. *et al.* Ultra-fast artificial neuron: generation of picosecond-duration spikes in a current-driven antiferromagnetic auto-oscillator. *Sci. Rep.* **8**, 15727 (2018).

60. Manipatruni, S., Nikonov, D. E. & Young, I. A. Beyond CMOS computing with spin and polarization. *Nat. Phys.* **14**, 338–343 (2018).

61. Manipatruni, S. *et al.* Scalable energy-efficient magnetoelectric spin–orbit logic. *Nature* **565**, 35–42 (2019).

62. Chanthbouala, A. *et al.* A ferroelectric memristor. *Nat. Mater.* **11**, 860–864 (2012).

63. Bur, A. *et al.* Electrical control of reversible and permanent magnetization reorientation for magnetoelectric memory devices. *Appl. Phys. Lett.* **98**, 262504 (2011).

64. Kolhatkar, G. *et al.* BiFe$_{1-x}$Cr$_x$O$_3$ Ferroelectric Tunnel Junctions for Neuromorphic Systems. *ACS Appl. Electron. Mater.* **1**, 828–835 (2019).

65. Chumak, A. V., Karenowska, A. D., Serga, A. A. & Hillebrands, B. Magnon spintronics. *Nat Mater* **11**, 1505–1549 (2015).

66. Brächer, T. & Pirro, P. An analog magnon adder for all-magnonic neurons. *J. Appl. Phys.* **124**, 152119 (2018).

67. Cherepov, S. *et al.* Electric-field-induced spin wave generation using multiferroic magnetoelectric cells. *Appl. Phys. Lett.* **104**, 082403 (2014).

68. Jaeger, H., Lukoševičius, M., Popovici, D. & Siewert, U. Optimization and applications of echo state networks with leaky- integrator neurons. *Neural Networks* **20**, 335–352 (2007).

69. Maass, W., Natschläger, T. & Markram, H. Real-time computing without stable states: a new framework for neural computation based on perturbations. *Neural Comput.* **14**, 2531–60 (2002).

70. Nomura, H. *et al.* Reservoir computing with dipole-coupled nanomagnets. *Jpn. J. Appl. Phys.* **58**, 070901 (2019).

71. Nakane, R., Tanaka, G. & Hirose, A. Reservoir Computing with Spin Waves Excited in a Garnet Film. *IEEE Access* **6**, 4462–4469 (2018).

72. Torrejon, J. *et al.* Neuromorphic computing with nanoscale spintronic oscillators. *Nature* **547**, 428–431 (2017).

73. Prychynenko, D. *et al.* Magnetic Skyrmion as a Nonlinear Resistive Element: A Potential Building Block for Reservoir Computing. *Phys. Rev. Appl.* **9**, 014034 (2018).

74. Bourianoff, G., Pinna, D., Sitte, M. & Everschor-Sitte, K. Potential implementation of reservoir computing models based on magnetic skyrmions. *AIP Adv.* **8**, 055602 (2018).

75. Alawad, M. & Lin, M. Survey of Stochastic-Based Computation Paradigms. *IEEE Trans. Emerg. Top. Comput.* **7**, 98–114 (2019).

76. Pinna, D. *et al.* Skyrmion Gas Manipulation for Probabilistic Computing. *Phys. Rev. Appl.* **9**, 064018 (2018).

77. Camsari, K. Y., Salahuddin, S. & Datta, S. Implementing p-bits With Embedded MTJ. *IEEE Electron Device Lett.* **38**, 1767–1770 (2017).

78. Camsari, K. Y., Faria, R., Sutton, B. M. & Datta, S. Stochastic p -Bits for Invertible Logic. *Phys. Rev. X* **7**, 031014 (2017).

79. Zand, R., Camsari, K. Y., Datta, S. & DeMara, R. F. Composable Probabilistic Inference Networks Using MRAM-based Stochastic Neurons. *ACM J. Emerg. Technol. Comput. Syst.* **15**, 1–22 (2019).

80. Camsari, K. Y., Chowdhury, S. & Datta, S. Scalable Emulation of Stoquastic Hamiltonians with Room Temperature p-bits. *arXiv 1810.07144* (2018).

81. Traversa, F. L. & Di Ventra, M. Polynomial-time solution of prime factorization and NP-complete problems with digital memcomputing machines. *Chaos An Interdiscip. J. Nonlinear Sci.* **27**, 023107 (2017).

82. Borders, W. A. *et al.* Integer factorization using stochastic magnetic tunnel junctions. *Nature* **573**, 390–393 (2019).

83. Traversa, F. L. & Di Ventra, M. Universal Memcomputing Machines. *IEEE Trans. Neural*

*Networks Learn. Syst.* **26**, 2702–2715 (2015).

84. Di Ventra, M. & Pershin, Y. V. The parallel approach. *Nat. Phys.* **9**, 200–202 (2013).

85. Cai, J., Fang, B., Wang, C. & Zeng, Z. Multilevel storage device based on domain-wall motion in a magnetic tunnel junction. *Appl. Phys. Lett.* **111**, 182410 (2017).

86. Fukami, S. & Ohno, H. Perspective: Spintronic synapse for artificial neural network. *J. Appl. Phys.* **124**, 151904 (2018).

87. Kampfrath, T. *et al.* Terahertz electrical writing speed in an antiferromagnetic memory. *Sci. Adv.* **4**, eaar3566 (2018).

88. Traversa, F. L. f2000 memcomputing solution. Available at: https://miplib.zib.de/instance_details_f2000.html.

89. Traversa, F. L. pythago7824 memcomputing solution. Available at: https://miplib.zib.de/instance_details_pythago7824.html.